\documentclass[
reprint,
prc,
aps,
floatfix,
]{revtex4-2}

\usepackage{graphicx}
\usepackage{bm}
\usepackage{xcolor}
\usepackage{physics}
\usepackage[hidelinks]{hyperref}

\begin{document}

\title{
Radiative strength functions from the energy-localized Brink-Axel hypothesis
}

\author{Oliver C. Gorton} 
\email{gorton3@llnl.gov}
\affiliation{Lawrence Livermore National Laboratory, P.O. Box 808, L-414,
Livermore, California 94551, USA}

\author{Konstantinos Kravvaris}
\affiliation{Lawrence Livermore National Laboratory, P.O. Box 808, L-414,
Livermore, California 94551, USA}

\author{Jutta E. Escher}
\affiliation{Lawrence Livermore National Laboratory, P.O. Box 808, L-414,
Livermore, California 94551, USA}

\author{Calvin W. Johnson}
\affiliation{San Diego State University,
San Diego, California 92182, USA}

\begin{abstract}
Radiative strength functions (RSFs) model the bulk electromagnetic response of
highly-excited nuclei and are critical inputs for statistical reaction codes. In
this paper, we present a definition of the RSF that is consistent with
Hauser-Feshbach reaction codes and that can be efficiently computed with the
shell model using the Lanczos strength-function (LSF) method. We introduce a
variant of the shell model LSF method that exploits the energy-localized
Brink-Axel hypothesis, which makes it possible to compute both electric and
magnetic RSFs across all energies relevant to capture reactions. We verify
agreement with the conventional definition of RSFs with benchmark calculations
of $^{24}$Mg, then present novel results for $^{56}$Fe. For $^{56}$Fe we find
that: (i) the M1 RSF shape evolves smoothly with excitation energy, consistent
with the energy-localized BrinklAxel hypothesis, (ii) both M1 and E1 transitions
contribute significantly to the radiative strength below the photo-absorption
threshold, and (iii) within the \textit{sdpf} model space, the strength below 3
MeV observed in Oslo-type experiments cannot be fully reproduced. These results
pave the way for a coherent microscopic description of RSFs and further
motivate the use of energy-dependent RSFs in modern reaction codes. 
\end{abstract}

\maketitle

\section{\label{sec:Intro}Introduction}

In 1952, \citet{blatt1991theoretical} commented ``an exact theoretical
description of [radiative transitions to or from highly excited levels] is
impossible, since the radiative transition probabilities can be calculated only
if the nuclear wave functions are known''. Today, this statement continues to
represents a compelling challenge since models for radiative decay are an
essential ingredient in the Hauser-Feshbach (HF) reaction
codes~\cite{koning2023talys, ormand2021monte, kawano2020coh3, herman2007empire,
young1992gnash} that underpin many applications in basic
science~\cite{mumpower2016impact, mumpower2017estimation,
arnould2020astronuclear} and nuclear technology~\cite{hayes2017applications,
kolos2022current}. Many theoretical and experimental efforts have sought to
understand how to approximate radiative decay~\cite{brink1955aspects,
axel1962electric, kopecky1990test, kopecky1993radiative, goriely2019simple},
resulting in phenomenological models called gamma (photon) strength
functions~\cite{kopecky1990test}. We prefer the nomenclature \textit{radiative}
strength functions (RSFs) to align with the specific form used in HF theory.

From an experimental standpoint, RSFs are notoriously difficult to measure
unambiguously across a wide energy range, especially for radioactive
isotopes~\cite{escher2012compoundnucleara, larsen2011analysis,
larsen2018enhanced, ingenberg2025nuclear}. Thus, the first RSF models
approximated the photo-emission probability using the photo-absorption cross
section of nuclear ground states~\cite{brink1955aspects}, which can be measured
in a more straightforward manner~\cite{berman1975measurements}. However, this
approximation runs afoul of neutron capture data which predicts nonzero strength
below the photo-absorption energy threshold. Low-energy photons are attributable
to the exponential growth of transitions available to highly excited
nuclei~\cite{schwengner2013lowenergy, frauendorf2015lemar} and models have long
included temperature-dependent corrections to account for these
effects~\cite{kadmenskii1983gamma, kopecky1990test, goriely2019simple}.  Because
this energy regime plays a crucial role in neutron-capture reactions during
nucleosynthesis~\cite{mumpower2016impact, mumpower2017estimation}, there have
been significant efforts to understand the low-energy behavior of these
functions~\cite{schwengner2010e1, schwengner2013lowenergy, brown2014large,
martini2016largescale, sieja2017electric, schwengner2017low,
karampagia2017lowenergy, midtbo2018consolidating, frauendorf2022evolution,
fanto2024lowenergy, chen2025origin}. However, the most common experimental
method cannot probe transitions below a few MeV~\cite{larsen2011analysis}. 

With advancements in many-body methods and computing hardware, there are now
practical methods for computing accurate nuclear wave functions from microscopic
models, enabling RSFs to be calculated and tabulated. While mean-field methods
like Hartree-Fock Bogoliubov (HFB) coupled with the quasiparticle random-phase
approximation (QRPA) have been broadly successful for this
task~\cite{goriely2018gognyhfb, li2024multipole, peru2025photon,
goriely2025qrpa, xu2026systematic}, the low-energy predictions of these models
are currently supplemented with empirical correction factors informed by
shell-model calculations~\cite{goriely2018gognyhfb}. Although, recent work has
shown it is possible to overcome this limitation~\cite{goriely2025qrpa}. 

Calculation of low-energy magnetic RSFs using the large-scale shell model (LSSM)
is now commonplace~\cite{frauendorf2022evolution, frauendorf2015lemar,
midtbo2018consolidating, larsen2018enhanced, brown2014large,
schwengner2013lowenergy, sieja2017electric, sieja2023brink,
karampagia2017lowenergy}, and some calculations even include electric
contributions (which require larger model spaces)~\cite{sieja2017electric,
liddick2019benchmarking, dahl2025microscopic}. LSSM calculations have been
instrumental to interpret the enhanced strength, typically referred to as the
low-energy enhancement (LEE)~\cite{brown2014large, naqvi2019nuclear,
larsen2018enhanced}, observed using the Oslo method~\cite{rekstad1983study,
schiller2000extraction}. However, a LSSM-only description remains impractical
due to computational costs driven by a) the apparent need to calculate thousands
of individual wave functions to capture the LEE, and b) the extended model
spaces required for the highly collective and cross-shell electric transitions.

Now that it is possible to predict the radiative decay of highly excited nuclei
from microscopic models, a new challenge arises: to predict RSFs across a wide
range of energies within a single framework while also building the underlying
nuclear wave functions from realistic nuclear interactions and accurate
many-body methods. We address this challenge by demonstrating that RSFs,
including the LEE, can be computed using the Lanczos strength-function method
(section~\ref{sec:method}), so long as the Lanczos pivot vector is constructed
from an excited state wave function. This approach becomes apparent only after
we clarify the relationship between traditional sum-rule strength functions and
the RSFs used in statistical reaction theory (section~\ref{sec:theory}). Using
this new approach, we explore the RSF of $^{56}$Fe and show that (i) the
magnetic dipole RSF shape evolves with excitation energy, consistent with an
energy-localized Brink-Axel hypothesis, (ii) electric dipoles transitions
contribute significantly to the low-energy radiative strength, and (iii) within
the \textit{sdpf} model space, the LEE observed in Oslo-type experiments cannot
be fully reproduced below 3~MeV (section~\ref{sec:results}).

\section{Theory}\label{sec:theory}

In this section, we review the multiple definitions of RSFs in the literature.
Then, we present a simplified definition which is compatible with HF
conventions while enabling a new computational technique.

\subsection{Preliminaries}\label{sec:prelim}

In HF theory~\cite{hauser1952inelastic}, the probability that an excited state
of the compound nucleus will decay with a photon of energy $E_\gamma$ is
proportional to the so-called transmission coefficient
$T^{\mathrm{X}L}(E_\gamma)$, with $X$ denoting the electromagnetic character of
the transition ($X=M$ for magnetic and $X=E$ for electric) and $L$ denoting the
multipolarity of the transition. The transmission coefficient is in turn
proportional to the  \textit{radiative strength function} (RSF; denoted
$f(E_\gamma)$) which is the energy-averaged radiative decay width, rescaled by a
phase-space factor of the photon energy $E_\gamma$~\cite{blatt1991theoretical}:
\begin{equation}
    T^{\mathrm{X}L}(E_\gamma) = 2\pi E_\gamma^{2L+1} 
    \overleftarrow{f}^{\mathrm{X}L}(E_\gamma).\label{eq:tran}
\end{equation} 
Here $E_\gamma = E_i - E_f$ is the transition energy between an initial level
$i$ with energy $E_i$ and a final level $f$ with energy $E_f$. The arrow
$\leftarrow$ indicates ``downward'' strength ($E_i > E_f$). As written in
Eq.~\eqref{eq:tran}, the RSF depends only on $\mathrm{X}L$ and $E_\gamma$. In a
many-body microscopic theory, however, other dependencies such as the structure
and quantum numbers of the initial and final levels may apply.

The standard definition of RSFs in terms of nuclear structure quantities was
given by Bartholomew in 1973~\cite{bartholomew1973gammaray}. As restated
by~\citet{kopecky2025low}, the Bartholomew definition of the RSF is the product
of an average decay width and a level-density-like quantity:
\begin{equation}\label{eq:bart0}
    \overleftarrow{f}^{\mathrm{X}L}_\text{Barth.}(E_\gamma) 
    = 
    \frac{1}{E_\gamma^{2L+1}}\langle \Gamma_{fi}^{\mathrm{X}L}\rangle_i\rho(E_i, j_i).
\end{equation}
Here, $\langle \Gamma_{fi}^{\mathrm{X}L}\rangle_i$ is the average partial decay
width between initial levels $i$ and a single final level $f$, for transitions
of multipole type $\mathrm{X}L$. The average is over all initial levels $i$ near
$E_i$ leading to transitions with energy $E_\gamma$ (and conserving angular
momentum). $\rho(E_i, j_i)$ is the density of levels near $E_i$ with spin $j_i$
which can decay to $f$.

The partial decay width for the level $i$ can be written in terms of reduced
transition probabilities ($B$-values):
\begin{equation}\label{eq:gamma}
    \Gamma_{f\leftarrow i}^{\mathrm{X}L} 
    = \frac{8\pi(L+1)}{ L[(2L+1)!!]^2}
    k_\gamma^{2L+1} 
    B^{\mathrm{X}L}_{f\leftarrow i},
\end{equation}
where $k_\gamma=E_\gamma/(\hbar c)$, and $B^{\mathrm{X}L}_{f\leftarrow i}$
directly relates to nuclear wave functions and transition operators:
\begin{equation}\label{eq:bvalues}
    B^{\mathrm{X}L}_{f\leftarrow i} 
    = \frac{1}{2j_i+1}|\langle \psi_f||\mathcal{M}^{\mathrm{X}L}||\psi_i \rangle|^2.
\end{equation}
The nuclear wave functions $\ket{\psi_x}$ have energy $E_x$, total angular
momentum-parity $j^{\pi_x}_x$, and magnetic substates $m_x$. The $B$-values are
reduced matrix elements (indicated by the double bar), meaning the final state
magnetic quantum numbers ($m_f$) have been summed and the initial substates
($m_i$) averaged~\cite{edmonds1957angular}. Matrix elements for the multipole
operator $\mathcal{M}^{\mathrm{X}L}$ can be found in standard
texts~\cite{blatt1991theoretical, brussard1977shellmodel, suhonen2007nucleons}.  

The contemporary method for computing RSFs with the LSSM was first established
by~\citet{schwengner2013lowenergy} and made explicit by~\citet{brown2014large}.
Inserting Eq.~\eqref{eq:gamma} into Eq.~\eqref{eq:bart0}, one can write down
a special form of Eq.~\eqref{eq:bart0} for magnetic
dipole transitions~\cite{frauendorf2022evolution, naqvi2019nuclear,
larsen2018enhanced, brown2014large, schwengner2013lowenergy}:
\begin{align}\label{eq:schwengner}
    \overleftarrow{f}^\text{M1}_\text{Barth.} = 
   \frac{16\pi}{9(\hbar c)^{3}} 
   [ B^\mathrm{M1}_{f\leftarrow i}]_{fi}(E_i, j^{\pi_i}_i, E_\gamma)
   \rho(E_i, j_i^{\pi_i}),
\end{align}
where $[\cdot ]_{fi}$ is the ``pixel-average'': all non-zero $B$-values are
sorted into $(100~\textrm{keV})^2$ ``pixels'', with $E_i$ along one axis and
$E_\gamma$ along the other, then divided by the number of $B$-values in each
pixel~\cite{brown2014large}. Finally, Eq.~\eqref{eq:schwengner} is averaged over
$j_i^{\pi_i}$ and $E_i$ when plotted as a function of
$E_\gamma$~\cite{frauendorf2022evolution, midtbo2018consolidating}. For the
remainder of the paper we omit $\pi_x$ labels, since for a given X$L$ transition
only a single value is allowed.

The details of the level density's spin and parity dependence in
Eq.~\eqref{eq:schwengner} has produced some confusion, as explained in the
appendix of Ref.~\cite{midtbo2018consolidating}. If we define RSFs primarily in
terms of discrete transition probabilities (the natural language of the shell
model), without reference to level densities, these issues disappear. 

When dealing with discrete transition probabilities, one typically works with
\textit{sum-rule strength functions} of the form~\cite{ring2004nuclear,
utsuno2015photonuclear, johnson2015systematics}:
\begin{equation}\label{eq:sumrulesf}
    S^{\mathrm{X}L}(E_i, E_\gamma) 
    = \sum_f \delta(E_f - E_i + E_\gamma) B^{\mathrm{X}L}_{f\leftarrow i},
\end{equation}
which do not explicitly refer to level densities. While Eq.~\eqref{eq:sumrulesf}
is directly proportional to the photo-absorption cross section when $E_i=0$ (and
$E_f > E_i$), it is incompatible with the conventions of HF theory (see
Appendix~\ref{sec:relations}). Thus, in order to reconcile the two, we present a
level-density-free (LDF) definition of the RSF which also simplifies the
algorithm for computing RSFs within the LSSM framework. 

\subsection{Level-density-free definition}

\begin{figure}[ht]
    \centering
    \includegraphics[width=0.65\linewidth]{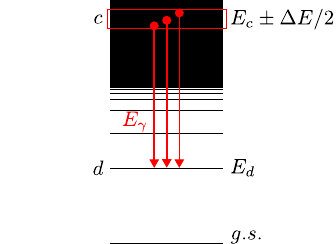}
    \caption{Depiction of the radiative strength function,
    Eq.~\eqref{eq:defgsf}. The red arrows indicate partial decay widths from
    levels $c \to d$; the box indicates an energy-average over partial decay widths
    from within $E_c \pm \Delta E/2$.}
    \label{fig:schematic}
\end{figure}

HF theory expresses all cross sections in the language of \textit{decay} widths,
regardless of whether a given nuclear state is absorbing or emitting radiation.
We therefore find it useful to replace the notation of ``initial levels'' $i$
and ``final levels'' $f$. Instead, we define ``compound-nucleus (CN) levels''
$c$ and ``de-excited levels'' $d$ to imply $E_c > E_d$, as shown in
Fig.~\ref{fig:schematic}. 

Consider a nuclear level $d$ impinged upon by an unpolarized beam of photons
with energy $E_\gamma$. The energy-averaged cross section for this level to be
excited to an energy near $E_c=E_d+E_\gamma$ is (\textit{c.f.}
Ref~\cite{thompson2009nuclear} p.324):
\begin{align}\label{eq:a}
    \langle \sigma^{\mathrm{X}L}_\gamma \rangle (E_\gamma) 
    &=
    \frac{\pi}{k_\gamma^2} 
    \sum_{j_c} \frac{(2j_c+1)}{w(s)(2j_d+1)} 
    T^{\mathrm{X}L}_{d,j_c}(E_\gamma),
\end{align}
which includes a sum over the spins $j_c$ of the CN levels, where $\bm j_c = \bm
j_d + \bm L$ (i.e., all spins allowed by angular momentum coupling).
$T_{d,j_c}^{\mathrm{X}L}(E_\gamma)$ are the photon transmission coefficients,
$w(s)=2$ is the statistical factor for incoming photons, and the remaining
statistical factors account for the fact that
$T_{d,j_c}^{\mathrm{X}L}(E_\gamma)$ describe \textit{decay} probabilities for
$c\to d$, even though Eq.~\eqref{eq:a} describes a physical process for $d\to
c$. By convention, transmission coefficients always refer to particle
\textit{emission} (in this case, photons).

The transmission coefficients are related by Eq.~\eqref{eq:tran} and the
Bartholomew definition Eq.~\eqref{eq:bart0} to the energy-averaged partial decay
widths for $c \to d$~\cite{bartholomew1973gammaray, blatt1991theoretical,
axel1962electric}:
\begin{align}\label{eq:bb}
    T^{\mathrm{X}L}_{d,j_c}(E_\gamma) 
    &= 2\pi 
    \langle \Gamma^{\mathrm{X}L}_{d\leftarrow c} \rangle_c 
    \rho(E_c, j_c).
\end{align}
At this point we can remove an approximation implicit in this expression. It can
be shown that Eq.~\eqref{eq:bb} can equally be written as an energy-average,
rather than the product of an average and a level density (c.f.
\citet{blatt1991theoretical}, p.653):
\begin{align}
\nonumber T_{d,j_c}^{\mathrm{X}L}(E_\gamma) 
    &= 2\pi \left (\frac{1}{N_c}
    \sum_{c'} 
    \delta_{j_{c'}j_c}
    \Gamma_{d\leftarrow c'}^{\mathrm{X}L}(E_\gamma) \right) 
    \left ( \frac{N_c}{\Delta E}\right ) \\
    &= 2\pi \frac{1}{\Delta E} 
    \sum_{c'} 
    \delta_{j_{c'}j_c}
    \Gamma_{d\leftarrow c'}^{\mathrm{X}L}(E_\gamma)\label{eq:d},
\end{align}
where $N_{c}=N_c(E_c, j_c)$ is the number of partial widths included in the sum
(the same number as in the average). Eq.~\eqref{eq:bb} is an approximation of
Eq.~\eqref{eq:d} because the number of compound levels which contribute to the
average, $N_c$, depends on the specific de-excited level $d$ and is merely
bounded by the level density: $N_c/\Delta E \leq \rho(E_c, j_c)$. While this
approximation is effective (see section~\ref{sec:validate}), we favor
Eq.~\eqref{eq:d} because it depends only on partial widths, the natural language
of the shell model, and enables the efficient computational techniques that we
introduce in section~\ref{sec:method}.

Finally, we combine Eq.~\eqref{eq:tran} with Eq.~\eqref{eq:d} to obtain the main
result of this section, a level-density-free (LDF) definition of the radiative
strength function:
\begin{align}\label{eq:defgsf}
    f^{\mathrm{X}L}_{d, j_c}(E_\gamma) = 
    \frac{1}{E_\gamma^{2L+1}} \frac{1}{\Delta E} 
    \sum_{c'} \delta_{j_{c'}j_c} 
    \Gamma^{\mathrm{X}L}_{d\leftarrow c'},
\end{align}
where we omit the arrow $\leftarrow$ since from here on we deal exclusively
with radiative strengths for which $E_c > E_d$ is implied (as in
Fig.~\ref{fig:schematic}). In this form, the RSF appears as the
``$E_\gamma$-gated'' total decay width to a de-excited level $d$ from levels
with angular momentum $j_c$. A key detail of Eq.~\eqref{eq:defgsf} is that,
regardless of whether the RSF is used to compute photo-absorption or radiative
decay, the sum over decay widths $\Gamma_{d\leftarrow c}$ always runs over CN
levels $c$. This detail not only sets the correct statistical factors in the
$B$-values (i.e. $1/(2j_c+1)$), but it also determines which level density the
RSF is proportional to ($\rho(E_c)$, not $\rho(E_d)$). In
Appendix~\ref{sec:relations}, we show the consistency of this convention in
relation to different concepts: capture cross sections in terms of $B$-values,
generalized transmission coefficients for radiative decay, and sum-rule strength
functions. 

In a microscopic theory, $f_{d,j_c}^{\mathrm{X}L}(E_\gamma)$ may depend on the
de-excited level $d$ and the spin $j_c$. In HF theory, however, RSFs are
generally assumed to be independent of both of these quantities. Reducing
Eq.~\eqref{eq:defgsf} to a spin-independent form is done with a simple average:
\begin{align}
    f^{\mathrm{X}L}_{d}(E_\gamma) 
    &= \langle f^{\mathrm{X}L}_{d,j_c}(E_\gamma) \rangle_{j_c} \\
    \label{eq:db}
    &= \frac{1}{n_{j_c}} \sum_{j_c} f^{\mathrm{X}L}_{d,j_c}(E_\gamma),
\end{align}
where $n_{j_c}$ is the number of $j_c$ allowed by angular momentum rules.
Furthermore, the strong Brink-Axel hypothesis holds that the RSF should be
independent of $d$. Historically, this hypothesis was used to justify using the
RSF of the ground state, $f^{\mathrm{X}L}_{d=1}(E_\gamma)$ ($E_{d=1}=0$), as the
RSF for decays to any other level~\cite{brink1955aspects, axel1962electric}.
However, it has been shown that the strong hypothesis is not supported from a
microscopic perspective, and at most one should assume the energy-localized
Brink-Axel (ELBA) hypothesis~\cite{johnson2015systematics, herrera2022modified}.

The work following \citet{schwengner2013lowenergy} has already taken the first
step to reject the strong Brink-Axel hypothesis, since it explicitly includes
radiative strength to excited states; this move is what led to the discovery
that the LSSM could reproduce the LEE (see Eq.~\eqref{eq:schwengner}
discussion). In our formalism, the equivalent energy- and spin-independent RSF
is:
\begin{equation}\label{eq:fsa}
    f^{\mathrm{X}L}_{(SA)}(E_\gamma) = 
    \langle f^{\mathrm{X}L}_{d, j_c}(E_\gamma) \rangle_{d, j_c},
\end{equation}
where the average is first over $j_c$ and second over $d$. Eq.~\eqref{eq:fsa}
can therefore be directly compared to the averaged version of
Eq.~\eqref{eq:schwengner}. In this paper, we take the next step towards RSFs
that reflect the ELBA hypothesis.

To summarize, the LDF-RSF should be interpreted as the energy-averaged radiative
strength to a specific nuclear level $d$, and the same form applies for HF
calculations of both photo-absorption and radiative decay. As discussed in the
next section, the LDF-RSF formulation allows us to use the Lanczos strength
function method to compute it.

\section{Method development}\label{sec:method}

In this section, we propose a new approach for computing RSFs that mirrors the
ELBA hypothesis and uses the Lanczos methods~\cite{caurier2005shell,
whitehead1980shellmodel, bloom1984gamow} developed for sum-rule strength
functions. In section~\ref{sec:validate}, we validate our LDF-RSF,
Eq.~\eqref{eq:defgsf}, using LSSM calculations of $^{24}$Mg in the \textit{sd}
valence space, for which thousands of converged wavefunctions can be obtained.
Then, in section~\ref{sec:elba}, we show that the low-energy behavior of the RSF
emerges as a result of the excitation energy of the de-excited level, rather
than the mere fact of including many transitions among excited states. Finally,
in section~\ref{sec:new}, we suggest an efficient approximation scheme for
radiative strength functions (RSFs) using the existing Lanczos strength function
method (LSF), which avoids the need to compute transitions between many excited
states.

\subsection{Tests in the \textit{sd}-shell}\label{sec:validate}

\begin{figure}[ht]
    \centering
    \includegraphics[width=\linewidth]{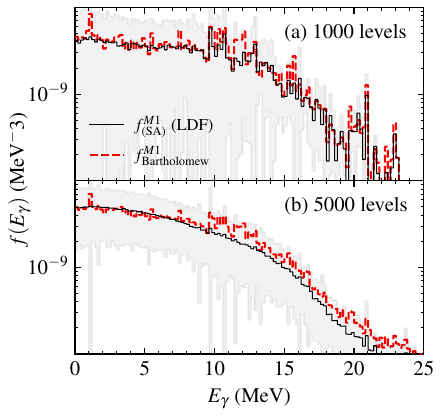}
    \caption{Agreement of the Bartholomew formula (dashed) and the present
    level-density-free (LDF) formula (solid), demonstrated for the M1 RSF of
    $^{24}$Mg. Panel a) was computed using the first 1000 levels and panel b)
    with the first 5000 levels (see text for details). All calculations used a
    bin size $\Delta E= 200$~keV. The gray band shows the standard deviation of
    the RSFs $f_d$ within each bin, which are averaged to obtain
    $f_\text{(SA)}^{\mathrm{X}L}$.}
    \label{fig:bart-vs-micro2}
\end{figure}

We find that our new LDF-RSF formula~\eqref{eq:defgsf} agrees overall with the
Bartholomew prescription of Eq.~(\ref{eq:bart0}), although there are small
deviations, due to binning effects (see Fig.~\ref{fig:bart-vs-micro2}). In the
Bartholomew formula as implemented with pixel-based averaging, strengths are
first simultaneously averaged over both CN levels $c$ and de-excited levels $d$,
then multiplied by the level density $\rho(E_c, j_c)$. This second step
approximately ``undoes'' the first average over CN levels $c$. As a result,
information is lost about the number of levels $c$ with nonzero transition
strengths to particular levels $d$, and it is assumed that this number is given
by $\rho(E_c, j_c)$ for all $d$. For the largest $E_\gamma$ values corresponding
to transitions to low-lying states, the level density $\rho(E_c, j_c)$ tends to
over-predict the actual number of nonzero transitions, since not all levels $c$
connect to all de-excited levels $d$. Our formulation of Eq.~\eqref{eq:defgsf}
avoids this unnecessary approximation. 

To perform the two RSF calculations in each panel of
Fig.~\ref{fig:bart-vs-micro2}, we used the same set of LSSM
$B(\mathrm{M1})$-values computed for $^{24}$Mg. We used the \textit{sd} model
space and the USDB~\cite{brown2006new} effective interaction, which we
diagonalized with the {\tt BIGSTICK} shell model
code~\cite{johnson2018bigstick}. We used the standard one-body M1 operator with
coupling constants $g_l^p = 1$, $g_l^n=0$, $g_s=5.5857$, and $g_s^n=-3.8263$. In
study (a), we use the lowest 1000 states of $^{24}$Mg which generates
180$\cdot10^3$ nonzero $M1$ transitions with $E_d<E_c$. In study (b), we use the
5000 lowest states of $^{24}$Mg which generates 4$\cdot10^6$ transitions. Both
(a) and (b) are plotted with a bin size of $\Delta E=200$~keV.

In addition to the averaged strength (which in the final step averages $f_d$
over all de-excited levels $d$), we also present the standard deviation of $f_d$
within each bin. The gray bands in Fig.~\ref{fig:bart-vs-micro2} show the +1 and
-1 standard deviations about the mean $\langle f_d\rangle$. This band does not
necessarily correspond to an uncertainty, but rather it shows the variability of
the radiative strength functions from one de-excited state to another (for one
thing, the distribution of strengths is not normal, but closer to a $\chi^2$
distribution). The size of the band is therefore related to the concept of
Porter-Thomas fluctuations of the radiative
strengths~\cite{porter1956fluctuations, campo2019test, markova2022nuclear}:
since the partial widths follow a $\chi^2$ distribution, the mean tends to
fluctuate over a broad range until a sufficient number of terms are summed (the
bands also therefore depend on the size of the bins). 

As a consequence, as we increase the number of states in the average (either by
increasing the number of levels in the calculation or increasing the bin size),
this standard deviation decreases. In the HF framework, the applicability of
radiative strength functions relies on the level density near the de-excited
state $\rho(E_d, j_d)$ being sufficiently high such that the Porter-Thomas
fluctuations are sufficiently small.

\subsection{Energy-localized Brink-Axel hypothesis}\label{sec:elba}

We can now demonstrate that the shape of the RSF evolves with the energy of the
de-excited level $d$ on which the RSF is built. While the ELBA hypothesis holds
that the total strength of a transition operator evolves smoothly with the
energy of $d$~\cite{johnson2015systematics}, we further show that is it
primarily the low-energy strength (i.e. below the photo-absorption threshold)
that increases with the energy of the de-excited level.

\begin{figure}[ht]
    \centering
    \includegraphics[width=\linewidth]{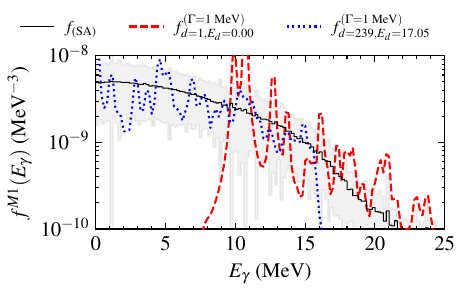}
    \caption{Contribution of two individual strength functions
    $f_d^\mathrm{M1}(E_\gamma)$ to the overall reduced strength
    $f_\mathrm{(SA)}^{\mathrm{X}L}(E_\gamma)$, showing that the ground state has
    a unique shape, while a random excited state has a partial radiative
    strength function with a shape compatible with the total reduced radiative
    strength function. The ground state ($d=1$) has no strength below about
    7~MeV due to a physical lack of states close enough in energy. The $d=239$
    strength, on the other hand, has significant strength approaching
    $E_\gamma=0$.}
    \label{fig:fdependence}
\end{figure}

Using Eq.~\eqref{eq:defgsf}, it is trivial to decompose the LDF-RSF into
contributions from individual de-excited levels $d$. We do so for the ground
state $d=1$ and a highly excited state $d=239$. For the RSFs to specific
de-excited states $f_d$, we take the additional step of folding with a
Lorentzian of width $\Gamma = 1$~MeV in order to smooth the visual
representation, as is sometimes done to approximately account for missing
collectivity at higher energies~\cite{utsuno2015photonuclear} and effects due to
the continuum~\cite{kruse2019nocore} (see Fig.~\ref{fig:fdependence}). 

For the ground state, there is no strength below $E_\gamma = 4$~MeV, since the
lowest $1^+$ state is predicted around this energy; we call this the
photo-absorption threshold. Overall, the shape of this RSF could be approximated
by a single Lorentzian function centered around 10~MeV, which is the standard
form for the M1 RSF in HF codes~\cite{capote2009ripl}. 

For the highly-excited level near $E_d=17.05$~MeV ($d=239$), however, maximum
strength occurs below 1~MeV. Notably, the sudden drop in the strength above
$16$~MeV in this case is not physical, but reflects the limited number of levels
computed in this example: of the 5000 states the maximum excitation energy is
33~MeV; with $E_d=17$~MeV, the maximum $E_\gamma = 33 - 17 = 16$~MeV. A scan
across all levels from $d=1$ to $d=239$ (not shown) shows a gradual saturation
of the M1 strength at low energies. The behavior exemplified by the $d=239$
level appears to be typical for highly-excited levels, and comparison to the
mean further suggests this is the norm.

The area under the two curves in Fig.~\ref{fig:fdependence} approximately
reveals what has already been explored by~\citet{johnson2015systematics}, which
is that the integrated strength functions (proportional to sum rules) violate
the strong Brink-Axel hypothesis. The M1 strength function built on the ground
state has overall smaller strength than that of the excited states. As a result,
the RSF built on the ground state is not representative of the average decay
strength. At first, this difference in total strength suggests one is doomed to
compute thousands of wave functions and millions of transition probabilities
whenever one wants to compute the RSFs relevant for capture reactions (which
always involved highly-excited nuclei). However, there are two key points that
mitigate this conclusion:

(i) \citet{johnson2015systematics} and~\citet{herrera2022modified} showed that
the total (sum-rule) strength of a level varies smoothly with the energy of that
level, leading to the ELBA hypothesis. The maximum range of this total strength
varies by a factor of 5-6 for M1 isovector and E2 isoscalar responses, but much
less for $E1$ response, which in part explains the success of the strong
Brink-Axel hypothesis for E1 strengths. Therefore, if we are interested in the
strength for levels in a certain energy range, it is reasonable to expect that
an energy-localized average should be representative. It is also known that
accurate strength functions for individual levels can be obtained using
approximate wave functions using the Lanczos strength-function
method~\cite{whitehead1980shellmodel, bloom1984gamow, caurier2005shell}, again
due to the ELBA hypothesis~\cite{johnson2020exact, herrera2022modified},

(ii) The present results suggest that RSFs built on a single level $d$ (which
are closely related to the energy-distribution of the total sum-rule)
tend to follow a predictable trend, so long as $E_d$ is in a region where the
level spacing is small compared to the $E_\gamma$ of interest. In other words,
the RSFs built on low-lying states seem to be the exception to a more universal
trend. This observation suggests we can learn a general low-energy behavior of
the RSF from a small number of excited states.

In the next subsection, we leverage these two considerations to introduce a new
scheme to compute RSFs.

\subsection{Lanczos strength function method}\label{sec:new}

Inspired by the ELBA hypothesis, we seek to approximate the RSF at low-energies
using a small number of individual levels $d$ positioned several MeV above the
ground state. The ELBA hypothesis suggests that the strength function should
vary slowly for levels nearby in energy. By averaging the individual strength
functions of a small number of nearby excited states, we should be able to
approximate the average RSF. 

Additionally, there are well established methods to rapidly compute sum-rule
strength functions for a single level, namely, the Lanczos strength function
method (LSF). The LSF method approximates the strength function of a single
level using a modified Lanczos algorithm~\cite{whitehead1980shellmodel,
bloom1984gamow, caurier2005shell}. Each iteration of the LSF method essentially
resolves a new moment of the strength distribution, rather than generating one
transition probability at a time. In this subsection, we demonstrate the
effectiveness of this method to approximate RSFs for a single level $d$. Here we
highlight: (i) a key difference from sum-rule strength function methods is that
we sum only radiative strengths ($E_c > E_d$), and (ii) we assume
spin-independence of the strength in order to simplify calculation of the matrix
elements. We give further details of our LSF implementation in
Appendix~\ref{sec:lanczos}.

\begin{figure}[ht]
    \centering
    \includegraphics[width=\linewidth]{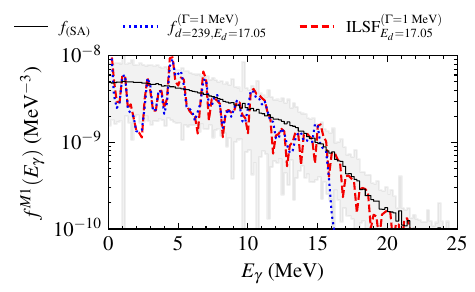}
    \caption{The radiative strength function for a single state can be computed
    accurately and to higher energy with the LSF method. We demonstrate with a
    level with $E_d=17.05$ near the neutron separation energy
    $S_n(^{24}\mathrm{Mg})=16.531$~MeV. We compare the result from the complete
    list of transitions against that from the ILSF method (see text). The state
    was selected to have $j_d=0$ so that Eq.~\eqref{eq:bjf0} applies,
    simplifying the calculation.} \label{fig:gsflt}
\end{figure}

The LSF method can approximate the RSF for individual de-excited levels $d$. We
show an example in Fig.~\ref{fig:gsflt}. In this case, we generated excited wave
functions (interior eigenstates) using thick-restart block
Lanczos~\cite{wu2000thick}, as done by \citet{herrera2022modified}.
Additionally, we specifically targeted pivot states with $j_d=0$ in order to
simplify the strength function calculation, as discussed in
Appendix~\ref{sec:lanczos}. We then applied the M1 operator and performed 200
Lanczos iterations on the pivot to obtain the reduced matrix elements for the
strength function calculation. We label this scheme the
\textit{interior-eigenstate Lanczos strength-function} (ILSF) method.

The error of the LSF method relative to the exact calculation is negligible
below the energies where states are available for the exact calculation
($E_\mathrm{max} - E_d \approx 33 - 17 = 16$~MeV). Above this energy, where the
exact method fails, the LSF method provides results closely following the trend
of the average strength, $f_\mathrm{(SA)}^{XL}$.

\begin{figure}[ht]
    \centering
    \includegraphics[width=\linewidth]{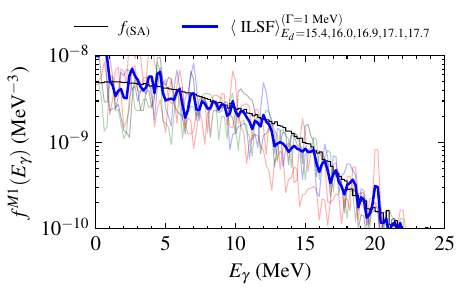}
    \caption{The Lanczos strength-function method with interior eigenvalues
    (ILSF) can be further improved by averaging the partial strength functions
    of several approximate excited levels. Here we use five $j_d=0$ interior
    eigenvectors with energies between 15 and 18 MeV as pivots. The
    fluctuations of individual strengths averages out to closely follow the
    average RSF $f_\mathrm{(SA)}^\mathrm{M1}$ computed with 5000 levels.}
    \label{fig:gsfltm}
\end{figure}

The LSF method by itself approximates only the strength function for a single
level, $f_d^{XL}$, and yields fine structures unique to the pivot level $d$.
To approximate the average radiative strength function, we average over five
nearby $j_d=0$ levels (obtained with an interior eigenstate method). The result
is shown in Fig.~\ref{fig:gsfltm}.

While the averaged ILSF result still contains local structures from its
constituent individual strength functions, we have significantly reduced the
computational cost compared to the full calculation: converging the 5000 levels
used in the full $f_\mathrm{(SA)}^{XL}$ required more than 18000 Lanczos
iterations (one \textit{iteration} is essentially one matrix-vector
multiplication). By contrast, for the ILSF method, approximate convergence of 5
interior wave functions required only 80 block iterations, followed by 200
iterations for each level to compute the strength distributions. In total, ILSF
used about 1400 iterations, reducing the cost by more than ten-fold.
Furthermore, by not having to store multiple vectors in memory/disk, the ILSF
approach scales significantly better in larger systems.

\section{Results for Iron}\label{sec:results}

We now turn to the computationally demanding $^{56}$Fe nucleus, for which
existing shell model methods to compute the E1 RSF up to energies relevant for
radiative capture calculations would be impractical.

From an experimental standpoint, $^{56}$Fe was one of the first nuclei observed
to have a low-energy enhancement (LEE) of the RSF. The $^{56}$Fe LEE has been
observed in $^{3}$He scattering by~\citet{voinov2004large}
(Fig.~\ref{fig:gsf-e1}: red dots) and~\citet{algin2008thermodynamic} (blue
triangles), and $(p,p'\gamma)$ scattering by~\citet{larsen2013evidence} (black
squares), along with the proton inelastic scattering data
from~\citet{jingo2018studies} (green crosses). However, there is currently no
microscopic model that can reproduce the magnitude of strength measured below
3~MeV. LSSM calculations indicate M1 transitions are not the source of the
surplus strength~\cite{midtbo2018consolidating, brown2014large}, and so far it
has been impractical to compute the low-energy E1 strength with the LSSM. Adding
to the tension between different approaches, a re-analysis of decay widths from
neutron-resonance data suggested significantly weaker LEE~\cite{kopecky2025low},
and \citet{kopecky2025low} argued that the scattering experiments should be
revisited.

In this section, we compare to the systematic model calculations of
\citet{goriely2018gognyhfb, goriely2019reference}, so we summarize the relevant
points from their work. The calculations of \cite{goriely2018gognyhfb} begin
with axially symmetric deformed quasiparticle random-phase approximation (QRPA)
based on the finite-range D1M Gogny force (to be denoted as ``D1M+QRPA''). The
E1 and M1 strengths obtained then require corrections due to well-characterized
systematic deficiencies. The strengths are modified in three ways. The strengths
are: (i) shifted down in $E_\gamma$ (ranging from -0.5~MeV at low energies to
-5~MeV at 21~MeV); (ii) folded with simple Lorentzians (E1 are given variable
widths $\Gamma = 7 - A/45$~MeV, M1 are given fixed widths $\Gamma=0.5$~MeV); and
(iii) combined with correction terms added to account for strength to excited
states:
\begin{align}\label{eq:d1mqrpa0lim}
    \overleftarrow{f}^\mathrm{E1}_\mathrm{QRPA+0lim}(E_\gamma) 
    &= f_\text{QRPA}^\mathrm{E1}(E_\gamma) + f_0U/[1+e^{(E_\gamma-\epsilon_0)}] \\
    \overleftarrow{f}^\mathrm{M1}_\text{QRPA+0lim}(E_\gamma) 
    &= f_\text{QRPA}^\mathrm{E1}(E_\gamma) + Ce^{-\eta E_\gamma}.
\end{align}
Here, $U$ is the excitation energy of the decay nucleus in MeV, and $f_0,
\epsilon_0, C$, and $\eta$ are free parameters adjusted to shell model
calculations and experimental data for radiative decay~\cite{goriely2018gognyhfb}.
The final modified calculations are denoted ``D1M+DRPA+0lim''.

\subsection{M1 strength in the \textit{pf-shell}}\label{sec:m1}

We used two schemes to compute the M1 RSF in the \textit{pf}-shell valence space
using the GX1A interaction~\cite{honma2004new}. The full configuration
interaction matrix dimension of $^{56}$Fe in this space is more than 500 million
(M-scheme). While converging a small number of levels is not difficult, it would
be impractical to compute and store the thousands of wave functions necessary to
cover the full energy range up to about 20~MeV (after which the strength is
negligible). We therefore apply the ILSF method and compare the results with
another approach, namely, a basis-truncation scheme that reduces the
computational cost of converging many levels.

\begin{figure}[ht]
    \centering
    \includegraphics[width=\linewidth]{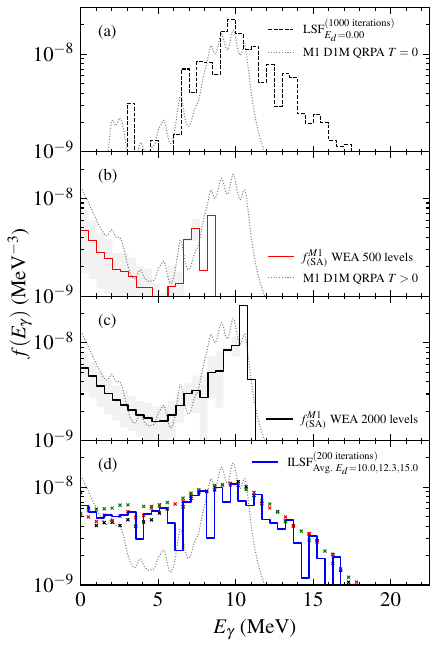}
    \caption{The M1 RSF for $^{56}$Fe for various energies of the de-excited
    level, computed within the \textit{pf} shell with the
    GX1A~\cite{honma2004new} interaction. a) The LSSM LSF strength to the ground
    state is comparable to the D1M QRPA calculation at zero temperature. Our
    result produces more strength at higher energies. b) With the weak
    entanglement approximation (WEA), we reproduce the qualitative shape of D1M
    QRPA $T>0$ calculation, which includes a phenomenological term fit to LSSM
    calculations. With only 500 levels, the calculation does not produce
    strength above 8.3~MeV. c) If we quadruple the number of WEA levels to 2000,
    we reach strength up to 10.7~MeV, including the dipole peak around 10~MeV.
    The distribution begins to flatten. d) We obtain the RSF up to 20~MeV by
    using the ILSF method introduced in the text, and averaging over partial
    RSFs for three levels above 10~MeV. Including higher excitations clearly
    leads to a flatter distribution of strength. Present calculations use
    $\Delta E=500$~keV. }
    \label{fig:gsf-m1}
\end{figure}

As with $^{24}$Mg, the shape of the $^{56}$Fe M1 RSF evolves with excitation
energy. As expected, the RSF built on the ground-state has no low-energy
strength (Fig.~\ref{fig:gsf-m1}a), while the RSF obtained when we average over
many de-excited levels shows an exponential LEE (Fig.~\ref{fig:gsf-m1}b).
Unexpectedly, we find that the evolution proceeds with even higher energies. By
quadrupling the number of excited levels included, we find that the slope of the
LEE decreases (Fig.~\ref{fig:gsf-m1}c). We can interpret this in light of the
results of section~\ref{sec:elba}, which showed that it is the energy of level
$d$ involved in the RSF, not the number of levels, that is responsible for the
low-energy strength. The flattening trend is reinforced if we extend our
calculations to even higher excitation energies (Fig.~\ref{fig:gsf-m1}d), made
possible by the ILSF method.

In the following, we discuss the four calculations in
Fig.~\ref{fig:gsf-m1} (a-d) which include progressively higher energies and
demonstrate the shape evolution of the M1 RSF .

(a) First, we used the LSF on the exactly converged ground state of our
\textit{pf} shell calculation. That is, we calculate
$f_{d=1}^\mathrm{M1}(E_\gamma)$. The overall shape converges after a few Lanczos
iterations. However, to avoid introducing uncertainty from the Lorentzian
folding procedure, we performed enough iterations (1000) such that the binned
distribution has only a few empty bins at a bin size of $\Delta E=500$~keV (see
Fig.~\ref{fig:gsf-m1}a). We find relatively good agreement with the
zero-temperature D1M+QRPA~\cite{goriely2018gognyhfb}. We find essentially the
same peak energy, without energy-shifting the strengths as was done for the QRPA
calculation~\cite{goriely2018gognyhfb}. Interestingly, our calculation produces
a long tail of strength above the peak. The strength around 3~MeV, which is
thought to be attributable to a scissors mode~\cite{heyde2010magnetic}, appears
in similar locations in both calculations.

(b) To compute a RSF averaged over many $^{56}$Fe levels $d$, we had to reduce
the computational cost of converging the wave functions. We used the
weak-entanglement approximation (WEA)~\cite{gorton2024weak, johnson2025weak}, an
importance truncation scheme that has been shown to perform well for RSF
calculations~\cite{gorton2023problem}. We give further details for the WEA
calculations in Appendix~\ref{sec:wea}. Using 500 WEA wave functions, we compute
the averaged radiative strength function $f_\text{(SA)}^{XL}$ over all 500 levels
(Fig.~\ref{fig:gsf-m1}b). The highest energy level is around 9.9~MeV, which sets the
upper-bound for photon energies. The resulting calculation is in agreement with
the D1M+QRPA+0lim $T>0$ shape~\cite{goriely2019reference, goriely2018gognyhfb},
which includes a phenomenological term inspired by LSSM calculations of M1
strengths (first proposed by~\citet{schwengner2013lowenergy}).

(c) By extending the previous calculation of the averaged RSF to use 2000 WEA
wave functions, three changes occur. The first is trivial: the RSF now reaches
$E_\gamma = 10.7$~MeV (Fig.~\ref{fig:gsf-m1}c). The second is that the standard
deviation of the individual strengths decreases, as shown by the gray bands.
This reduction can be interpreted in terms of Porter-Thomas fluctuations: as the
nucleus reaches higher excitation energies, more transitions contribute to
each $E_\gamma$ bin, bringing the distribution closer to the
mean~\cite{axel1962electric}. The third is that the LEE shape begins to flatten.

(d) The damping of the LEE slope (and increase in total strength) continues at
higher excitation energies, as shown in Fig.~\ref{fig:gsf-m1}d. Here, we
converged three interior eigenstates (wave functions) at 10.0, 12.3, and
15.0~MeV using thick-restart block-Lanczos. Then, using the ILSF method
previously discussed, we computed the average strength assuming the ELBA
hypothesis. Using the ILSF method, we can predict the RSF above
$E_\gamma=20$~MeV. The highest photon energies involved with these transitions
correspond to transitions from levels around $E_c = 30$~MeV; it would
not be computationally possible to converge the number of individual levels to
reach these energies with the brute-force method use above (even with the WEA).
One might be concerned that states at these energies may not be well described
in the \textit{pf}-space alone. However, we repeated these calculations in the
\textit{sdpf} valence space and found the results comparable.

The flattening of the M1 strength with excitation energy can be interpreted in
view of earlier results. \citet{karampagia2017lowenergy} showed the exponential
LEE in shell model calculations is a feature of transitions constrained to a
single orbit; as higher orbits are mixed into the model space, the strength is
``quenched'', leading to a flatter distribution. \citet{midtbo2018consolidating}
also showed that the M1 strength flattens as one moves away from a closed shell,
which can be understood to follow from the same mixing of orbitals. For our
results, mixing of valence space orbitals increases with excitation energy,
rather than artificial model space restrictions or proximity to closed shells.

From the systematics of strength function sum
rules~\cite{johnson2015systematics}, one should already anticipate that the
total M1 strength of excited states should be larger than that of the ground
state. What is novel about the present findings is an understanding of where (in
photon energies) that extra strength appears, and the additional step of
translating those strengths into reaction theory inputs.

We have shown that the M1 RSF strength distribution evolves slowly with the
excitation energy of the levels involved. Despite this new subtlety, we arrive
at the same conclusion from earlier LSSM calculations~\cite{brown2014large,
midtbo2018consolidating, karampagia2017lowenergy}, which is that the M1 RSF
cannot explain the LEE extracted from Oslo-type data. However, due to
computational costs, no calculation of the E1-contributions to the LEE has yet
been made for $^{56}$Fe. We address this need in the next subsection.

\subsection{E1 strength in the \textit{sdpf}-shell}

To compute the E1 strengths, one must include, at a minimum, two or more single
particle orbits in the model space with opposite parity and a difference in
angular momentum of one unit. To meet this minimum we use the \textit{sdpf}
shell model space ($1d_{5/2}, 2s_{1/2}, 1d_{3/2}, 1f_{7/2}, 2p_{3/2}, 1f_{5/2},
2p_{1/2}$) and the SDPFMU-DB interaction~\cite{iwata2016largescale,
utsuno2012shape}. It is likely that the $1g_{9/2}$ orbital plays a role in the
GDR of $^{56}$Fe, but there is not presently a robust interaction and model
space publicly available to handle its inclusion. Inclusion of the $1g_{9/2}$
orbital may also influence the low-energy behavior of the RSF. This question
should be investigated once a suitable interaction and model space become
available.

The LSF on its own is insufficient to compute the RSF of $^{56}$Fe in the
\textit{sdpf} space, since the untruncated basis dimension ($2\times 10^{15}$)
is too large to obtain even a single wave function. To reduce the basis to a
manageable size, we used a truncation scheme based on energy centroids with the
{\tt tracer} code~\cite{johnson2025centroids} and the {\tt BIGSTICK} shell model
code~\cite{johnson2018bigstick}. As in, Ref.~\cite{gorton2024weak}, we assign to
each orbital $a$ an integer weight $w_a$. Each many-body configuration is given
a total weight $W$ which is the sum of the weights of the occupied orbitals. The
$M$-scheme space is truncated by keeping all configurations with weights up to a
maximum $W_\mathrm{max}$, defined relative to the minimum in the space
$W_\mathrm{min}$: $W_\mathrm{ex} = W_\mathrm{max} - W_\mathrm{min}$. We assigned
the following weights $w$ to each of the orbitals: $w(1d_{5/2}) = 1$;
$w(2s_{1/2}) = 2$; $w(1d_{3/2}) = 0$; $w(1f_{7/2}) = 3$; $w(2p_{3/2}) = 4$;
$w(1f_{5/2}) = 5$; $w(2p_{1/2}) = 6$. This choice of weights approximates a
truncation based upon the centroids (average energies) of orbital
configurations~\cite{horoi1994truncation,jiao2014correlatedbasis}. With
$W_\mathrm{ex}=11$, $^{56}$Fe has an $M$-scheme dimension of 520~million. 

\begin{figure}[ht]
    \centering
    \includegraphics[width=\linewidth]{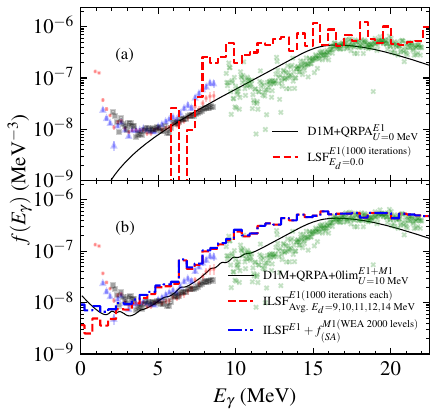}
    \caption{The LEE observed in several experiments cannot be explained by a
    combination of radiative M1 and E1 strength with the present
    \textit{sdpf}-valence-space calculations. We show (a) shell model
    calculations of the E1 RSF involving only the ground state of $^{56}$Fe
    using our LSF method, and (b) the excited-state E1 RSF generated with our
    ILSF method using five levels from 9~MeV to 14~MeV. The data shown are from
    $^{3}$He scattering by~\citet{voinov2004large} (red dots) and
    \citet{algin2008thermodynamic} (blue triangles), $(p,p'\gamma)$ scattering
by \citet{larsen2013evidence} (black squares), and proton-inelastic, isovector
giant dipole resonance data from~\citet{jingo2018studies} (green crosses). }
    \label{fig:gsf-e1}
\end{figure}

With the above approximations, we find the following results. The E1 RSF built
on the ground state with the LSF method has no strength below about
$E_\gamma=6$~MeV, which is where the first $1^-$ level occurs
(Fig~\ref{fig:gsf-e1}a). This is the qualitative result one would expect to
observe from E1 photo-absorption on the ground state of $^{56}$Fe. 

To calculate the RSF to the ground state, we
first used standard Lanczos to converge the $^{56}$Fe ground state wave
function. We performed no folding with Lorentzians or systematic shifting of
strength. The strength averaging was performed with a bin width of $\Delta
E=500$~keV. To compute the E1 transitions, we used standard one-body operators
with effective charges $e_p = +N/A = 30/56 = 0.536$ and $e_n = -Z/A = -26/56 =
-0.464$. The harmonic oscillator basis length $b=\sqrt{\hbar/(m\omega)}$ was set
to the Blomqvist and Molinari prescription~\cite{blomqvist1968collective}: $ b^2
= 41.467/( 45A^{-1/3} - 25A^{-2/3})\ \mathrm{fm}^2, $ which for $^{56}$Fe gives
$b=2.03$~fm. To remove spurious center of mass motion, we included a Lawson
term~\cite{gloeckner1974spurious} in all of our \textit{sdpf} calculations, both
while converging the pivot wave functions and when applying the LSF method. We
verified our resulting wave functions have COM energies less than 0.001~MeV,
which is negligible compared to their roughly 500~MeV relative binding energies.

Next, when we include higher excitation energies in the RSF calculation,
strength appears below the photo-absorption threshold ($\approx$ 6~MeV) which
was not seen in the RSF to the ground state (Fig.~\ref{fig:gsf-e1}b).
Calculation of this E1 LEE was only made possible with the ILSF method if we
first applied the same basis-truncation used to obtain the \textit{sdpf} ground
state.

To apply the ILSF method, we partially converged five $j_d=0$ levels near the
neutron separation energy $S_n(^{56}\text{Fe})=11.197$~MeV. As in the
\textit{pf}-space calculations, our wave functions were obtained with
thick-restart Lanczos, with the additional $cJ^2$ term to push-out $J>0$ levels.
We performed a total of 800 iterations, which yielded five levels within
0.004$\hbar$ of $J=0\hbar$, and excitation energies $E = 9.05,
10.2,11.3,12.5,13.7$~MeV. After applying the same E1 operator as for the ground
state RSF, we applied 1000 Lanczos iterations to each interior pivot.

Summing together the E1 strength from ILSF and the M1 strength from the WEA
calculation with 2000 levels, we present the total RSF for $^{56}$Fe computed in
the \textit{sdpf} model space. We chose to use this particular set of M1
strengths because it aligns with the excitation energies accessed in the Oslo
measurements (up to around the neutron separation energy). However, the
incompatibility with the Oslo data persists in all of our M1 calculations. The
E1 strength dominates the low-energy RSF above 3~MeV, below which the M1
strength dominates. Within the present \textit{sdpf} model space and
interaction, the shell model calculation cannot reproduce the RSF extracted from
Oslo data below 3 MeV. The remaining discrepancy below 3~MeV is between a factor
of 3 to 10.

We find a GDR energy centroid similar to that of the D1M+QRPA result
($\sim$16~MeV). Both of these microscopic predictions are low relative to the
experimental value around 18.6~MeV, and low relative to the systematic estimate
which predicts a double peak at 17.08 and 20.12~MeV~\cite{capote2009ripl}. Much
of the shell model strength is shifted lower in energy, seen as surplus strength
between 8 and 15~MeV. An isospin decomposition reveals this energy shift is
split by isospin-conserving transitions ($\Delta T=0$) which peak at 15~MeV, and
the isospin-changing transitions ($\Delta T=1$) peak near 22~MeV.

\section{Conclusion}

We have shown that a level-density-free definition of the radiative strength
function formula can simplify calculations, clarify interpretation in terms of
sum rules, and enable the LSF method to be applied. The LSF method essentially
resolves the moments of the radiative strength distribution, rather than
generating all individual transition probabilities, which is the current
practice. We showed one can approximate the full RSF with just a few Lanczos
pivots, owing to the ELBA hypothesis.

The present results reveal that the M1 strength of $^{56}$Fe evolves smoothly as
a function of excitation energy. Strength to the ground state only
(corresponding to a photo-absorption measurement) has the expected single-GDR
shape. An average over strength to many excited states up to 10~MeV reveals the
expected exponential LEE. However, as we move to higher energies, the RSF shape
flattens to a nearly constant distribution below the M1 GDR. Smooth evolution of
total strength with excitation energy is expected from the ELBA
hypothesis~\cite{johnson2015systematics, herrera2022modified}. The
shape-evolution of the strength \textit{distribution} follows as a logical
extension of the ELBA hypothesis, and can be understood in analogy to the known
correlation with nearness to a closed shell~\cite{midtbo2018consolidating} or
artificial model space restrictions~\cite{karampagia2017lowenergy}.

Using our novel ILSF approach, we also presented the first LSSM description of
the radiative strength function of $^{56}$Fe, including M1 and E1 contributions
from $E_\gamma=0$ to $20$~MeV . The results suggest that the \textit{sdpf}
valence space, lacking the $g_{9/2}$ orbital, is insufficient to accurately
describe the E1 GDR response of $^{56}$Fe, though future calculations will help
in clarifying the situation. The discrepancy relative to the Oslo-type
measurements below 3~MeV also remains unresolved. We leave development of a
robust model space and effective interaction including the $g_{9/2}$ orbital to
future work. For the purposes of constraining neutron capture rates, it will
also be important to include uncertainty quantification based on the underlying
effective interaction similar to~\cite{gorton2025toward}.

RSFs were originally developed assuming the strong Brink-Axel hypothesis so that
radiative decay could be constrained by photo-absorption experiments. As
microscopic models become favored for their predictive power, this restriction
no longer serves its use. Our results and the ILSF method presented here instead
further support the energy-localized Brink-Axel hypothesis, indicating that RSFs
should evolve with the energy of the de-excited level. Looking forward, we
suggest that current HF codes could be improved by accepting tabulated RSFs
which depend on both $E_\gamma$ and $E_d$, the energy of the de-excited level.

\begin{acknowledgments}
This work was performed under the auspices of the U.S. Department of Energy by
Lawrence Livermore National Laboratory under Contract DE-AC52-07NA27344, with
support from Laboratory Directed Research and Development project
No.~24-ERD-023. This material is also based upon work by CWJ supported by the
U.S. Department of Energy, Office of Science, Office of Nuclear Physics, under
Award Number  DE-FG02-03ER41272, and work done by OCG supported by the LLNL WPD
ACT award program. We thank the broader RETRO collaboration for providing useful
feedback.
\end{acknowledgments}

\appendix

\section{LDF-RSF in relation to other quantities}\label{sec:relations}

We show how our RSF, Eq.~\eqref{eq:defgsf}, relates to: sum-rule strength
functions, the sum-rule expression for photo-absorption, and generalized
transmission coefficients.

\subsection{Relation to sum-rule strength functions}

Following the discussion in Section~\ref{sec:theory}, one can further show that
the energy-differential, time-reversed version of the RSF, Eq.~\eqref{eq:defgsf}
has the same form as the sum-rule strength function, Eq.~\eqref{eq:sumrulesf}.
That is, if we suppose that the RSF is an energy-average of some differential
strength $d f_d^{XL}/dE_c$, then we can write:
\begin{equation}
    f^{XL}_{d}(E_\gamma) = 
     \frac{1}{\Delta E} 
     \int_{E_c-\Delta E/2}^{E_c+\Delta E/2} 
     \frac{d f_d^{XL}}{dE_c}(E_\gamma) dE_c,
\end{equation}
where we are not selecting a particular spin $j_c$ as in
Eq.~\eqref{eq:defgsf}. It follows that the differential strength function must
be:
\begin{equation}\label{eq:diffgsf}
    \frac{d f_d^{XL}}{dE_c}
    (E_\gamma)
    =
    \sum_{c} \delta(E_c - E_d + E_\gamma) 
    \frac{\Gamma^{XL}_{d\leftarrow c}}{E_\gamma^{2L+1}}.
\end{equation}
Eq.~\eqref{eq:diffgsf} and Eq.~\eqref{eq:sumrulesf} have similar forms in terms
of $B$-values. The key differences are: (i) the RSF contains additional
constants from in the definition of $\Gamma^{XL}_{dc}$, (ii) the RSF is often
given for specific spins $j_c$, and (iii) the RSF always sums over the index $c$
with $E_c > E_d$, whereas the sum-rule SF sums over all $f \neq i$; when
$S^{XL}$ is built on the ground state, it is proportional to the
absorption cross-section, but $d f_d^{XL}/dE_c$ contains the time-reversed
statistical factors and must be combined with the factors in Eq.~\eqref{eq:a}. Finally, because the RSF does not sum over all levels in
the model space (only those with $E_c > E_d$), the RSF does not strictly obey
sum rules.

\subsection{Relation to other expressions for capture}

As noted in \citet{blatt1991theoretical} (p. 392), HF theory is constructed such
that the absorption cross section is proportional to the probability for the
decay of the compound nucleus, hence the $B$-values for $c \to d$ in the
definition of the RSF, Eq.~\eqref{eq:defgsf}. However, the shell-model
expression for photo-absorption is typically written as~\cite{ring2004nuclear,
utsuno2015photonuclear, kruse2019nocore}:
\begin{equation}\label{eq:smcs}
    \sigma^{XL}_\gamma (E_\gamma)
    =
    \frac{8\pi^3(L+1)k_\gamma^{2L-1}}{ L[(2L+1)!!]^2} S^{XL}(E_d, E_\gamma),
\end{equation}
where $S^{XL}(E_d, E_\gamma)$ is given by Eq.~\eqref{eq:sumrulesf} (with
$B$-values for $d \to c$, summed over $c$). Clearly, the RSF~\eqref{eq:defgsf}
is not time-reversible ($\overleftarrow{f} \neq \overrightarrow{f}$) since
$B^{XL}_{c\leftarrow d} \neq B^{XL}_{d\leftarrow c}$. So, how do we reconcile
these two expressions for photo-absorption?

As a consistency check, we fully expand the photo-absorption cross section
Eq.~\eqref{eq:a} to show it is equivalent to Eq.~\eqref{eq:smcs}. We substitute
the transmission coefficients, Eq.\eqref{eq:d}, into the HF cross section
Eq.~\eqref{eq:a} and combine the sums over $j_c$ and $c$:
\begin{align}
    \langle \sigma^{XL}_\gamma \rangle (E_\gamma)
    &=
    \frac{\pi^2}{k_\gamma^2} 
    \sum_{j_c} \frac{(2j_c+1)}{(2j_d+1)} 
    \frac{1}{\Delta E} 
    \sum_{c'} 
    \delta_{j_{c'}j_c}
    \Gamma_{d\leftarrow c'}^{XL}(E_\gamma) \\
    &= 
    \frac{\pi^2}{k_\gamma^2} \frac{1}{\Delta E}
    \sum_{c} \frac{(2j_c+1)}{(2j_d+1)}
    \Gamma_{d\leftarrow c}^{XL}(E_\gamma).
\end{align}
Next, we expand the widths in terms of squared matrix-elements. 
After cancellation of statistical factors we have:
\begin{align}
    \langle \sigma^{XL}_\gamma \rangle (E_\gamma)
    &=
    \frac{8\pi^3(L+1)k_\gamma^{2L-1}}{ L[(2L+1)!!]^2}
    \frac{1}{\Delta E}\sum_{c} 
    \frac{1}{2j_d+1} \nonumber\\&\ \times
    \sum_{m_cm_dM} 
    \left | \langle j_d m_d | \mathcal{M}^{XLM} | j_c m_c \rangle \right |^2
\end{align}
Using the definition of $B$-values and the symmetry
$
    \left | \langle j_d m_d | \mathcal{M}^{XLM} | j_c m_c \rangle \right |^2 
    = \left | \langle j_c m_c | \mathcal{M}^{XLM} | j_d m_d \rangle \right |^2,
$
we obtain a cross section containing the $B$-values for $d \to
c$:
\begin{equation}
    \langle \sigma^{XL}_\gamma \rangle (E_\gamma)
    =
    \frac{8\pi^3(L+1)k_\gamma^{2L-1}}{ L[(2L+1)!!]^2}
    \frac{1}{\Delta E}\sum_{c} 
    B^{XL}_{c \leftarrow d},\label{eq:f}
\end{equation}
which is the energy-average of Eq.~\eqref{eq:smcs}.

\subsection{Relation to generalized transmission coefficients}

In the case of photo-absorption, the RSF Eq.~\eqref{eq:defgsf} should be
computed for a single level $|j_d\rangle$, the level absorbing photons. But what
about for radiative decay? In the case of decay from a bin $(E_c, j_c, \pi_c)$
to another de-excited bin $(E_d, j_d, \pi_d)$, a HF code computes the
\textit{generalized transmission coefficient} as:
\begin{align}\label{eq:gent}
    \mathcal{T}^{XL}(E_\gamma, E_c) 
    &= \int_{E_d-\Delta E/2}^{E_d+\Delta E/2} 
    T_{d}^{XL}(E_\gamma) \rho(E_d) dE_d,
\end{align}
where the spin and parity selection rules must be obeyed, but are not explicitly
shown. No additional statistical factor is necessary, since, by construction,
the reduced transition probabilities $B^{XL}_{d \leftarrow c}$ contained in
Eq.~\eqref{eq:defgsf} contain the correct factors already: a sum over de-excited
magnetic substates $m_d$ and an average over CN magnetic substates $m_c$.

\section{LSF implementation}\label{sec:lanczos}

We use the {\tt BIGSTICK} shell model code to apply the Lanczos
strength-function (LSF) method~\cite{johnson2018bigstick}. For a more detailed
description of the LSF method, we recommend the appendix of
\citet{herrera2022modified}. Here, we give a cursory introduction while
highlighting key details relevant for computing radiative strength functions.

The LSF method uses a modified Lanczos algorithm to compute the matrix elements
needed to compute radiative strength functions, given by sums over $|\langle
\psi_c | \mathcal{M}^{XL} | \psi_d \rangle|^2$ for an electromagnetic operator
$\mathcal{M}^{XL}$. The trick is to first create a Lanczos pivot vector by
applying the transition operator to the initial wave function: $| v_0 \rangle =
\mathcal{M}^{XL} | \psi_d \rangle$. Then, the Lanczos algorithm is applied,
generating approximate eigenstates $|\tilde \psi \rangle$ of the Hamiltonian.
For each $|\tilde \psi \rangle$, one computes the squared overlap with the the
pivot: $|\langle \tilde \psi_c | v_0 \rangle|^2$, which are the required matrix
elements. Because we energy-average over $c$ to obtain the RSF, the fact that
the Lanczos eigenstates $| \tilde \psi_c \rangle$ are only approximately
converged (and thus a superposition of nearby levels) becomes
unimportant~\cite{herrera2022modified, johnson2020exact}.

The radiative strength function to a particular de-excited state $d$ has the
form:
\begin{align}\label{eq:fexpl}
f_d^{XL}(E_\gamma) 
    &= 
    C^{XL}
    \frac{1}{\Delta E}
    \sum_{c}
    B^{XL}_{dc},
\end{align}
(we omit spin selection for brevity) where,
\begin{equation}\label{eq:cconst}
    C^{XL} = \frac{8\pi(L+1)}{ L[(2L+1)!!]^2}
    \left ( \frac{1}{\hbar c} \right ) ^{2L+1}.
\end{equation}
Because the $B$-values, Eq.~\eqref{eq:bvalues}, are defined with a sum over all
``final'' magnetic states and an average over ``initial'' magnetic states, the
order $d < c$ is important. The $B$-values are computed within a single-$M$ basis
using the Wigner-Eckart theorem:
\begin{align}
    B^{XL}_{dc} &= 
    \frac{1}{2j_c+1}|\langle j_d||\mathcal{M}^{XL}||j_c\rangle|^2 \\
    &=\frac{2j_d + 1}{2j_c+1} 
    \left | \frac{\langle j_d m | \mathcal{M}^{XL0} 
    | j_c m \rangle }{(j_c m L 0 | j_d m) } \right |^2\label{eq:bfromlt}.
\end{align}
By choosing the Lanczos pivot to be an excited state, $\mathcal{M}^{XL} | j_d
m_d \rangle$, the LSF algorithm yields the unreduced amplitudes: $|\langle
\tilde \psi_c | \mathcal{M}^{XL} | j_d m_d \rangle|^2$. However, in an $M=0$
basis, certain vector coupling coefficients are zero due to symmetries ($(j_i 0
j_f 0 | L 0)=0$ when $j_i+j_f-L$ is odd), which results in ``missing'' $L=1$
strengths when $j_i=j_f$ and $j_i, j_f$ are integers, which must be recovered
(although, we find their effect is negligible).

\begin{figure}[ht]
    \centering
    \includegraphics[width=\linewidth]{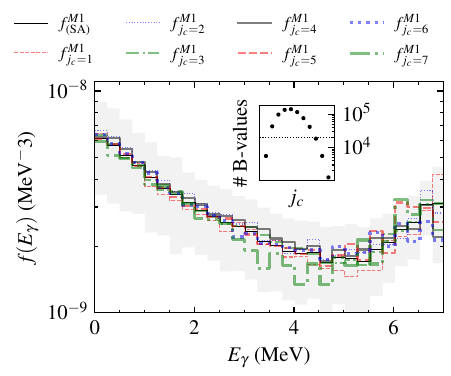}
    \caption{The RSF $f_{j_c}$ is independent of the spin of the decaying levels
    $j_c$, as illustrated by plotting the RSF from Fig.~\ref{fig:gsf-m1}(c)
    for different $j_c$. Deviations of specific $f_{j_c}$ arise when too few
    $B$-values contribute, which occurs when there are a limited number of levels
    with that spin included in the calculation. The inset shows the number of
    $B$-values used in each $f_{j_c}$ for $j_c=0,1,...,10\hbar$. RSFs computed
    from few $B$-values are susceptible to Porter-Thomas fluctuations (see
    section~\ref{sec:validate}), so we only plot $f_{j_c}$ with more than
    $2\cdot 10^4$ contributing $B$-values, excluding $j_c = 0, 8, 9, 10\hbar$.}
    \label{fig:spindependence}
\end{figure}

For an arbitrary choice of the pivot vector spin $j_d$, the states $\tilde
\psi_c$ produced by the Lanczos iterations will have a superposition of multiple
$j_c$, since $\mathcal{M}^{XL}$ is not a scalar operator. This makes it
difficult to compute the vector coupling constants in Eq.~\eqref{eq:bfromlt};
one must then first project out states of good $J$. However, if we choose a
$j_d=0$ wave function, only $j_c=1$ pivots are produced. Using the relation
$(j_1 0 j_2 0|00)^2=\delta_{j_1,j_2}/(2j_1+1)$, one can further show that for
$j_d=0$:
\begin{align}\label{eq:bjf0}
    B^{XL}_{dc}(j_d=0) 
    &= \left | \langle j_d m | \mathcal{M}^{XL} | j_c m \rangle \right |^2\\
    &= \left | \langle j_c m | \mathcal{M}^{XL} | j_d m \rangle \right |^2.
\end{align}
Therefore, using any $j_d=0$ wave function to create the pivot greatly
simplifies the calculation. This restriction should be acceptable since: (i) HF
theory already assumes spin-independent RSFs, (ii) past shell model studies have
found minimal dependence on spin and parity~\cite{brown2014large,
karampagia2017lowenergy}, and (iii) our WEA calculations support the same
conclusion (Fig.~\ref{fig:spindependence}). Finally, the LSF RSF is computed in
an $M=0$ basis as:
\begin{align}\label{eq:lsfgsf}
f_{d (j_d=0)}^{XL}(E_\gamma) 
    &= 
    C^{XL}
    \frac{1}{\Delta E}
    \sum_{c}
    \left | \langle j_c 0 | \mathcal{M}^{XL} | j_d 0 \rangle \right |^2.
\end{align}
The spin-selected version is recovered by inserting $\delta_{j_c,j_c'}$.

\section{Constants and units}\label{sec:units}

For convenience, we provide numerical values for the constant,
Eq.~\eqref{eq:cconst}, in order to produce RSFs with units of MeV$^{-(2L+1)}$
when the units of the $B$-values are absorbed into the constant. For $EL$
transitions, the $B$-values had units of $e^2(\text{fm})^{2L}$, where $\text{fm}$
is taken to be unity. With $e^2 = \alpha \hbar c$ (with CGS units, $1=4\pi
\epsilon_0$) where $\hbar c = 197.327$~MeV~fm and the dimensionless
fine-structure constant $\alpha = 1/137.036$:
\begin{align}
    C^\mathrm{E1} &= \frac{16\pi}{ 9}
    \left ( \frac{1}{\hbar c} \right ) ^{3} e^2(\mathrm{fm})^{2}
    = 
    1.04669 \cdot 10^{-6} \text{ MeV}^{-2},\\
    C^{E2} &= 
    \frac{24\pi}{30} \left ( \frac{1}{\hbar c} \right ) ^{5} e^2(\text{fm})^{4}
    = 5.83384 \cdot 10^{-12} \text{ MeV}^{-4}.
\end{align}
For $ML$ transitions, the $B$-values had units of $\mu_N^2(\text{fm})^{2L-2}$;
$\mu_N$ is nuclear magneton which in CGS units is $\mu_N = 0.105154 e\text{
fm}$. With these constants, 
\begin{equation}
    C^\text{M1} = \frac{16\pi}{ 9}
    \left ( \frac{1}{\hbar c} \right ) ^{3} \mu_N^2
    = 
    1.15737\cdot 10^{-8}\text{ MeV}^{-2}.
\end{equation}

\section{Weak entanglement approximation}\label{sec:wea}

An alternative method to reduce the computational cost of computing shell model
radiative strength functions is to use an appropriate many-body truncation to
reduce the model space. For this purpose, we use the weak entanglement
approximation (WEA)~\cite{gorton2024weak, johnson2025weak} to solve for the
lowest 500 wave functions, and then the lowest 2000 wave functions. The WEA is
an importance truncation method that selects proton and neutron basis components
in the limit of zero proton-neutron entanglement. It has been shown that the
method performs well for M1 RSF calculations~\cite{gorton2023problem}. We use a
WEA basis constructed with 10-percent of the proton and neutron subspaces, which
yields model dimensions reduced by a factor of 1000. (The largest J-scheme
dimension we diagonalize, for $J=5$, is $5.3 \cdot 10^{5}$). Using this scheme,
our WEA ground state energy is within 100~keV (.05\%) of the untruncated value,
which is comparable to typical uncertainties originating from the effective
interaction~\cite{brown2006new, gorton2025toward}. We separately converged the
lowest 500 levels for each $J$ from $J=0$ to $J=10$ and retain 500 or 2000
levels with the lowest energies.

\bibliographystyle{apsrev4-2}
\bibliography{library}

\end{document}